\documentclass[a4paper,11pt]{amsart}
\usepackage{amssymb}
\usepackage{amsmath}
\usepackage{amsfonts}
\usepackage{latexsym}
\usepackage[all]{xypic}
\usepackage{enumitem}
\usepackage{graphicx}
\usepackage{color}

\newcommand{\be}{\begin{equation*}}
\newcommand{\ee}{\end{equation*}}
\newcommand{\ben}{\begin{equation}}
\newcommand{\een}{\end{equation}}
\newcommand{\beqa}{\begin{eqnarray*}}
\newcommand{\eeqa}{\end{eqnarray*}}
\newcommand{\beqan}{\begin{eqnarray}}
\newcommand{\eeqan}{\end{eqnarray}}
\newcommand{\nn}{\nonumber}

\def\R{\mathbb{R}}

\def\cK{\mathcal{K}}
\def\Sym{\mathrm{Sym}}
\def\rM{\mathrm{M}}
\def\Aut{\mathrm{Aut}}
\def\Stab{\mathrm{Stab}}

\def\Iso{\mathrm{Iso}}

\def\fh{\mathfrak{h}}
\def\vu{\vec{u}}
\def\rGamma{\mathrm{\Gamma}}

\newcommand{\pd}{\partial}
\def\dd{\mathrm{d}}

\newcommand{\id}{\mathrm{id}}

\newcommand{\tr}{\mathrm{tr}}

\newcommand{\sign}{\mathrm{sign}}

\def\cC{\mathcal{C}}

\def\cG{\mathcal{G}}
\def\cH{\mathcal{H}}

\def\cM{\mathcal{M}}
\def\cN{\mathcal{N}}
\def\cP{\mathcal{P}}

\def\cX{\mathcal{X}}
\def\cR{\mathcal{R}}
\def\cU{\mathcal{U}}

\def\bLambda{\boldsymbol{\Lambda}}

\def\mD{\mathbb{D}}

\def\rD{\mathrm{D}}

\def\rS{\mathrm{S}}

\def\hv{\hat{v}}

\def\SO{\mathrm{SO}}
\def\ISO{\mathrm{ISO}}

\def\v0{{\vec{0}}}
\def\cT{\mathcal{T}}

\newcommand{\eqdef}{\stackrel{{\rm def.}}{=}}

\DeclareMathOperator{\arctanh}{arctanh}

\def\div{\mathrm{div}}
\def\grad{\mathrm{grad}}
\def\Hess{\mathrm{Hess}}
\def\hHess{\widehat{\Hess}}
\def\End{\mathrm{End}}
\def\Hom{\mathrm{Hom}}

\def\Crit{\mathrm{Crit}}
\def\dist{\mathrm{dist}}

\font\tencyr=wncyr10
\font\sevencyr=wncyr7
\font\fivecyr=wncyr5
\newfam\cyrfam

\textfont\cyrfam=\tencyr
\scriptfont\cyrfam=\sevencyr
\scriptscriptfont\cyrfam=\fivecyr

\newtheorem{thm}{Theorem}[section]

\newtheorem{prop}[thm]{Proposition}

\newtheorem{cor}[thm]{Corollary}

\theoremstyle{definition}
\newtheorem{definition}[thm]{Definition}

\newtheorem{rmk}[thm]{Remark}

\newcommand{\threepartdef}[6]
{
	\left\{
	\begin{array}{ll}
		#1 & \mbox{if} #2 \\
		#3 & \mbox{if} #4 \\
		#5 & \mbox{if} #6
	\end{array}
	\right.
}

\def\fN{\mathfrak{N}}

\def\rk{\mathrm{rk}}

\begin{document}

\title[Hesse manifolds and Hessian symmetries of cosmological
models]{Hesse manifolds and Hessian symmetries of multifield
cosmological models}

\author{Calin Iuliu Lazaroiu} \address{Horia Hulubei National
Institute of Physics and Nuclear Engineering (IFIN-HH),
Bucharest-Magurele, Romania} \email{lcalin@theory.nipne.ro}

\begin{abstract}
I give a brief overview of the mathematical theory of Noether
symmetries of multifield cosmological models, which decompose
naturally into {\em visible} and {\em Hessian} (a.k.a. `hidden')
symmetries. While visible symmetries correspond to those infinitesimal
isometries of the Riemannian target space of the scalar field map
which preserve the scalar potential, Hessian symmetries have a much
deeper theory. The latter correspond to {\em Hesse functions}, defined
as solutions of the so-called {\em Hesse equation} of the target
space. By definition, a {\em Hesse manifold} is a Riemannian manifold
which admits nontrivial Hesse functions -- not to be confused with a
Hess{\em ian} manifold (the latter being a Riemannian manifold whose
metric is locally the Hessian of a function). All Hesse $n$-manifolds
$\cM$ are non-compact and characterized by their {\em index}, defined
as the dimension of the space of Hesse functions, which carries a
natural symmetric bilinear pairing. The Hesse index is bounded from
above by $n+1$ and, when the metric is complete, this bound is
attained iff $\cM$ is a Poincar\'e ball, in which case the space of
Hesse functions identifies with $\R^{1,n}$ through an isomorphism
constructed from the Weierstrass map. More generally, any elementary
hyperbolic space form is a complete Hesse manifold and any Hesse
manifold whose local Hesse index is maximal is hyperbolic. In
particular, the class of complete Hesse surfaces coincides with that
of elementary hyperbolic surfaces and hence any such surface is
isometric with the Poincar\'e disk, the hyperbolic punctured disk or a
hyperbolic annulus. Thus Hesse manifolds generalize hyperbolic
manifolds. On a complete Hesse manifold $(\cM,G)$, the value of any
Hesse function $\Lambda$ can be expressed though the distance from a
{\em characteristic subset} of $\cM$ determined by
$\Lambda$. Moreover, the gradient flow of $\Lambda$ can be described
using the distance function.
\vspace{0.5cm}

Keywords: Riemannian geometry, Noether symmetries, cosmology. 

\vspace{0.5cm}

AMS subject classifications: 53-XX, 70-XX, 83F05
\end{abstract}

\maketitle

\section{Introduction and physics motivation}

Cosmological models with at least two real scalar fields are of
increasing interest in theoretical physics. In our previous work
\cite{hidden,Noether,genalpha,elem,modular}, we initiated a geometric
study of the classical dynamics of multifield cosmological models with
arbitrary scalar manifold (which we approached from a mathematically
rigorous perspective), summarizing some of our results in
\cite{Tim19,QTS,Nis}. Cosmological models with $n$ real scalar fields
and standard kinetic term are parameterized by a so-called {\em scalar
triple} $(\cM,\cG,V)$, where $\cM$ is a connected smooth manifold
(generally non-compact and of non-trivial topology) which is the
target space of the scalar field map, $\cG$ is a Riemannian metric on
$\cM$ which specifies the kinetic term of the scalar fields and $V$ is
a smooth real-valued function defined on $\cM$, which specifies the
scalar potential. Such models arise naturally in string theory, where
$(\cM,\cG)$ appears as a moduli space of string compactifications and
$V$ is induced by a flux on the compactification manifold or by
quantum effects. The classical cosmological model parameterized by
$(\cM,\cG,V)$ involves the {\em scale factor} $a\in
\cC^\infty(\R,\R_{>0})$ of a simply-connected
Friedmann-Lemaitre-Robertson-Walker spacetime and a smooth curve
$\varphi:\R\rightarrow \cM$ (whose parameter $t\in \R$ is called {\em
cosmological time}) subject to a system of ODEs known as the {\em
cosmological equations}:
\beqan
\label{ELred}
&&3H^2+2\dot{H}+\frac{1}{2} ||\dot{\varphi}||^2_\cG - V\circ \varphi=0\nn\\
&& (\nabla_t +3H)\dot{\varphi} + (\grad_\cG V)\circ \varphi=0~~,\\
&& \frac{1}{2}||\dot{\varphi}||_\cG^2+V\circ \varphi=3 H^2~~,\nn
\eeqan
where the dot indicates derivation with respect to $t$ and $H\eqdef
\frac{\dot{a}}{a}\in \cC^\infty(\R)$ is the {\em Hubble parameter}.
The last equation in this system is called the {\em Friedmann
equation}.  Notice that $a$ enters this system only through its
logarithmic derivative $H$. When $H$ is positive, eliminating
it through the Friedmann equation allows one to reduce \eqref{ELred}
to the single second order autonomous ODE:
\ben
\label{eomsingle}
\nabla_t \dot{\varphi}(t)+\sqrt{\frac{3}{2}} \left[||\dot{\varphi}(t)||_\cG^2+2V(\varphi(t))\right]^{1/2}\dot{\varphi}(t)+ (\grad_{\cG} V)(\varphi(t))=0~~,
\een
which defines a dissipative geometric dynamical system (in the sense
of \cite{Palais}) on the total space of the tangent bundle of
$\cM$. In general, little is known about the deeper behavior of this
dynamical system, some aspects of which were explored in
\cite{genalpha,elem,modular} and summarized in \cite{Tim19,QTS,Nis}.

Let $\cN\eqdef \R_{>0}\times \cM$ be the {\em configuration space} of
the variables $a$ and $\varphi$. The cosmological equations \eqref{ELred}
can be derived from the variational principle of the so-called {\em
minisuperspace Lagrangian}\footnote{The term ``minisuperspace'' is
historically motivated and has nothing to do with supersymmetry.}
$L_{\cM,\cG,V}:T\cN\rightarrow \R$, which is given by:
\beqan 
\label{L}
L_{\cM,\cG,V}(a,\dot{a},\varphi,\dot{\varphi})&\!\eqdef\!& - 3 a \dot{a}^2 + a^3
\left[\frac{1}{2} ||\dot{\varphi}||^2_\cG - V (\varphi)\right]~~,
\eeqan
supplemented by the {\em Friedmann constraint}:
\ben
\label{F}
\frac{1}{2}||\dot{\varphi}||_\cG^2+V(\varphi)=3 H^2~~.
\een
Here we identity $T\cN$ with the first jet bundle of $\cN$ and we abuse
notation as common in jet bundle theory. Notice that the Friedmann
constraint is non-holonomic.

The constrained Lagrangian description given by \eqref{L} and
\eqref{F} allows for systematic study of the Lie symmetries (see
\cite{Olver}) of \eqref{eomsingle} using the Noether method. In
\cite{hidden}, we exploited this point of view to classify those
cosmological models with $\dim\cM=2$ which admit Noether symmetries,
making the technical assumption that the metric $\cG$ is rotationally
invariant. As already pointed out in that reference, the latter
assumption is purely technical and not needed for the results of
loc. cit. Moreover, the approach of \cite{hidden} generalizes to
cosmological models parameterized by arbitrary scalar triples
$(\cM,\cG,V)$, leading to a deep mathematical theory. This
generalization is discussed in detail in the preprints \cite{Hesse}
and \cite{nscalar}. We summarize some of its results below, focusing
on those aspects which are most relevant to Riemannian geometers. For
notational simplicity, we rescale the physics-motivated scalar
manifold metric $\cG$ to:
\be
G\eqdef \frac{3}{8}\cG~~,
\ee
thus replacing $(\cM,\cG)$ by the {\em rescaled scalar manifold}
$(\cM,G)$ and $(\cM,\cG,V)$ by the {\em rescaled scalar triple}
$(\cM,G,V)$.

\

\paragraph{\bf Notations and conventions}

Throughout this paper, $\cM$ will denote a smooth, paracompact,
Hausdorff and connected $n$-manifold (which need not be
compact). The differential of
a function $f\in \cC^\infty(\cM)$ is denoted by $\dd f\in
\Omega^1(\cM)$, while its value at a point $m\in \cM$ is denoted by:
\be
\dd_m f=(\dd f)(m)\in T_m^\ast \cM=\Hom_\R(T_m\cM,\R)~~.
\ee
We use the notations:
\be
Z(f)\eqdef \{m\in \cM~\vert~f(m)=0\}~~,~~\Crit(f)\eqdef \{m\in \cM~~\vert~\dd_m f=0\}
\ee
for the zero and critical locus of $f$ and:
\be
\cM_f(a)\eqdef f^{-1}(\{a\})=\{m\in \cM~\vert~f(m)=a\}
\ee
for the level set of $f$ at $a\in \R$. We will often use the following
two operators determined by a Riemannian metric $G$ on $\cM$:

\begin{itemize}
\itemsep 0.0em
\item The {\em Killing operator} of $(\cM,G)$, defined as the
$\R$-linear first-order differential operator
$\cK_G:\cX(\cM)\rightarrow \rGamma(\cM,\Sym^2(T^\ast \cM))$ which
associates to any vector field $X\in \cX(\cM)\eqdef \Gamma(\cM,T\cM)$
the symmetrization of the covariant derivative of the 1-form
$X^\flat\in \Omega^1(\cM)$:
\be
\cK(X)\eqdef \Sym^2[\nabla(X^\flat)]~.
\ee
In local coordinates on $\cM$, we have:
\be
\cK_G(X)_{ij}\eqdef \frac{1}{2}\left[\nabla_i X_j+\nabla_jX_i\right]
=\frac{1}{2}\left(\partial_i X_j+\partial_jX_i-2\rGamma_{ij}^k X_k \right)~~,
\ee
where $\rGamma_{ij}^k$ are the Christoffel symbols:
\be
\rGamma_{ij}^k=G^{lk}\rGamma_{ijl}=\frac{1}{2}G^{lk}(\partial_j G_{il}+\partial_i G_{jl} 
- \partial_l G_{ij})~~
\ee
and we use implicit summation over repeated indices.
\item The {\em Hessian operator} of $(\cM,G)$, defined as the
$\R$-linear second order differential operator
$\Hess_G:\cC^\infty(\cM)\rightarrow \rGamma(\cM,\Sym^2 (T^\ast \cM))$
which associates to a smooth real-valued function $f$ defined on $\cM$
its Hessian tensor:
\be
\Hess_G(f)\eqdef \nabla \dd f.
\ee
In local coordinates on $\cM$, we have:
\be
\Hess_G(f)_{ij}\eqdef \Hess_G(f)(\pd_i,\pd_j)=\partial_i\partial_j f- \rGamma_{ij}^k \partial_k f~~.
\ee
\end{itemize}

\noindent Notice the relation:
\be
\cK_G(\grad_G f)=\Hess_G(f)~~ \, ~~\forall f\in \cC^\infty(\cM)~~.
\ee

\section{Noether symmetries of multifield cosmological models}

Let $(\cM,G)$ be a Riemannian manifold and $V\in \cC^\infty(\cM)$ be a smooth
real-valued function defined on $\cM$. Let $\cN\eqdef \R_{>0}\times \cM$.
We have a natural decomposition $T\cN = T_{(1)}\cN\oplus T_{(2)}\cN$,
where $T_{(1)}\cN$ and $T_{(2)}\cN$ are the pullbacks of the tangent bundles
$T \R_{>0}$ and $T \cM$ through the canonical projections
$p_1:\cN\rightarrow \R_{>0}$ and $p_2:\cN\rightarrow \cM$:
\be
T_{(1)} \cN\eqdef p_1^\ast(T\R_{>0})~~,~~T_{(2)} \cN\eqdef  p_2^\ast(T\cM)~~.
\ee
Hence any vector field $X \in\cX(\cN )$ decomposes as: $X =
X_{(1)} + X_{(2)}$, with $X_{(i)}\in \Gamma(\cN,T_{(i)}\cN)$. In local
coordinates $(U,a,\varphi^i)$ on $\cN$, we have:
\be
X_{(1)}(a,\varphi)=X^a(a,\varphi) \frac{\pd}{\partial
a}~~,~\quad~X_{(2)}(a,\varphi)=X^i(a,\varphi) \frac{\pd}{\pd
\varphi^i}~~,
\ee
where $X^a, X^i\in \cC^\infty(U)$ and $i=1,...,n$.

\subsection{The characteristic system of variational symmetries}

The following result reduces the study of Noether symmetries of the
minisuperspace Lagrangian to that of certain real-valued functions and
vector fields defined on $\cM$.

\begin{thm}{\rm \cite{nscalar}}
A vector field $X\in \cX(\cN)$ is a time-independent Noether symmetry
of the minisuperspace Lagrangian of the classical cosmological model
parameterized by the rescaled scalar manifold $(\cM,G)$ and by the scalar
potential $V$ iff it has the form:
\be
\label{Xsol}
X:=X_{\Lambda,Y}=\frac{\Lambda}{\sqrt{a}}\pd_a + Y- \frac{4}{a^{3/2}} (\grad_G \Lambda)~~,
\ee
where $\Lambda\in \cC^\infty(\cM)$ and $Y\in \cX(\cM)$ satisfy the
{\em characteristic system} of $(\cM,G,V)$:
\beqan
\label{LambdaS}
\Hess_G(\Lambda)=G \Lambda~~&,&~~\langle \dd V, \dd \Lambda \rangle_G= 2 V \Lambda \\
\label{YS}
\cK_G(Y)=0~~&,&~~Y(V)=0~~.
\eeqan
\end{thm}

\

\noindent Notice that the two equations above containing $\Lambda$
decouple from those containing $Y$, so the characteristic system
consists of two independent systems of linear PDEs: the {\em
$\Lambda$-system} \eqref{LambdaS} and the {\em $Y$-system} \eqref{YS}
of $(\cM,G,V)$. In local coordinates, the characteristic system reads:
\beqa
\label{Sindex}
\left(\partial_i \partial_j -\Gamma_{ij}^k \partial_k\right)
\Lambda=G_{ij}\Lambda~~ &,&~~G^{ij} \partial_i V \partial_j\Lambda =2V
\Lambda~~~~\\ \nabla_i Y_j+\nabla_j Y_i=0 ~~&,&~~Y^i \partial_i
V=0~~~,\nn
\eeqa
where we use Einstein summation over repeated indices $i,j,k=1,...,n$.

The solutions of the $Y$-system coincide with those Killing vector
fields $Y$ of $(\cM,G)$ which satisfy $L_YV=0$, i.e. with
infinitesimal isometries of $(\cM,G)$ which preserve the scalar
potential $V$. Such solutions form the Lie algebra of the {\em group
of symmetries} of the rescaled scalar triple $(\cM,G,V)$, defined as the
stabilizer of $V$ inside the group $\Iso(\cM,G)$ of isometries of
$(\cM,G)$:
\be
\Aut(\cM,G,V)\eqdef \{\psi\in \Iso(\cM,G)~\vert~V\circ \psi=V\}~~.
\ee
Notice that $\Aut(\cM,G,V)$ is a Lie group since it is a closed
subgroup of $\Iso(\cM,G)$.  For a generic triple $(\cM,G,V)$, we have
$\Aut(\cM,G,V)=1$, hence the $Y$-system of a generic rescaled scalar
triple admits only the trivial solution $Y=0$.

The first equation of the $\Lambda$-system will be called the {\em
Hesse equation} of $(\cM,G)$. The second equation of that system
(which we call the {\em $\Lambda$-$V$ equation}) can be solved
explicitly {\em for V} once we pick a solution of the first.

\begin{thm}
\label{thm:VLambda}
Let $\Lambda$ be a nontrivial solution of the Hesse equation.  Then
any smooth solution of the $\Lambda$-$V$-equation of $(\cM,G)$ takes
the form:
\ben
\label{Vsol}
V=\Omega||\dd \Lambda||^2_G=\Omega \left[\Lambda^2-(\Lambda,\Lambda)_G\right]~~,
\een
where $\Omega\in \cC^\infty(\cM\setminus \Crit(\Lambda))$ is an
arbitrary smooth function which is constant along the gradient flow of
$\Lambda$:
\ben
\label{lv0}
\langle \dd \Omega, \dd \Lambda\rangle_G=0~~.
\een
\end{thm}

\noindent For generic $(\cM,G)$, the Hesse equation admits only the trivial
solution $\Lambda=0$, which satisfies the $\Lambda$-$V$ equation with
any $V$.

The observations above imply, as expected, that a generic multifield
cosmological model has no Noether symmetries. Those special models
which {\em do} admit such symmetries are of particular interest in
theoretical physics.

\begin{definition}
A time-independent Noether symmetry $X=X_{\Lambda,Y}$ is called:
\begin{itemize}
\item {\em visible} if $\Lambda=0$, in which case $X=X_{0,Y}=Y$.
\item {\em Hessian} if $Y=0$, in which case
$X=X_{\Lambda,0}=\frac{\Lambda}{\sqrt{a}}\pd_a - \frac{4}{a^{3/2}} (\grad_G
\Lambda)$.
\end{itemize}
The rescaled scalar triple $(\cM,G,V)$ and corresponding cosmological
model are called {\em visibly-symmetric} or {\em Hessian} if they
admit visible or Hessian symmetries, respectively.
\end{definition}

\noindent Let $\fN_h(\cM, G,V)$, $\fN_v(\cM, G,V)$ and $\fN(\cM, G,V)$
be the linear spaces of Hessian, visible and time-independent Noether
symmetries. Then there exists an obvious linear isomorphism:
\be
\fN(\cM, G,V) \simeq_\R  \fN_h(\cM, G,V)\oplus \fN_v(\cM, G,V)~.
\ee

\begin{definition}
The cosmological model defined by the rescaled scalar triple
$(\cM,G,V)$ is called {\em weakly Hessian} if the Hesse equation of
$(\cM,G)$ admits nontrivial solutions. It is called {\em Hessian} if
$\fN_h(\cM,G,V)\neq 0$.
\end{definition}

\noindent Theorem \ref{thm:VLambda} implies:

\begin{cor}
The cosmological model defined by the rescaled scalar triple $(\cM,G,V)$ is
Hessian iff it is weakly Hessian and the scalar potential $V$ has the
form \eqref{Vsol}, with $\Omega$ a solution of \eqref{lv0}.
\end{cor}

Since the study of visible symmetries reduces to a classical problem
in Riemannian geometry, the mathematically interesting problem is to
classify all Hessian scalar triples and hence all Hessian
cosmological models. By the results above this reduces in turn to the
problem of characterizing those Riemannian manifolds whose Hesse
equation admits nontrivial solutions. Below, we describe a
few results in this direction, whose proof can be found in \cite{Hesse}.

\section{Hesse functions and Hesse manifolds}

Let us start by formulating the mathematical problem without reference
to its origin in physics.

\begin{definition}
Let $(\cM,G)$ be a Riemannian manifold of positive dimension. A {\em
Hesse function} of $(\cM,G)$ is a smooth solution $\Lambda\in
\cC^\infty(\cM)$ of the following linear second order PDE, which is
called the {\em Hesse equation} of $(\cM,G)$:
\ben
\label{HessCond}
\Hess_G(\Lambda)=G \Lambda
\een
and whose space of solutions we denote by $\cH_G(\cM)$.  The {\em Hesse
index} of $(\cM,G)$ is defined through:
\be
\fh_G(\cM)\eqdef \dim_\R \cH_G(\cM)~~.
\ee
The Riemannian manifold $(\cM,G)$ is called a {\em Hesse manifold} if
$\fh_G(\cM)>0$, i.e. if $\cH_G(\cM)\neq 0$.
\end{definition}

\begin{rmk}
The notion of Hesse manifold should not be confused with
that of Hess{\em ian} manifold, which means a Riemannian manifold
whose metric is given locally by the Hessian of a function.
\end{rmk}

\noindent We start by studying the Hesse equation.

\subsection{Relation to Hessian equations. Non-compactness of Hesse manifolds.}

The Hesse equation \eqref{HessCond} of $(\cM,G)$ is equivalent with a system of
so-called {\em Hessian equations} (see \cite{hessian1,hessian2}), namely
a Hessian system which includes both the Helmholtz and Monge-Amp\`ere equations of
$(\cM,G)$.

For any $f\in \cC^\infty(\cM)$ and $m\in \cM$, let:
\be
Q_m^G(f)(z)\eqdef \det \left[z \, \id_{T_m\cM}-\hHess_G(f)(m)\right]=\sum_{k=0}^n (-1)^k c_k^G(f)(m) z^{n-k}\in \R[z] 
\ee
be the characteristic polynomial of the $G_m$-symmetric linear
operator $\hHess_G(f)(m)$ $\in \End_\R(T_m\cM)$ obtained by raising an
index of the symmetric tensor $\Hess_G(f)(m)$, where $z$ is a formal
variable. The characteristic coefficients $c_k^G(f)(m)$ define smooth functions
$c_k^G(f)\in \cC^\infty(\cM)$ as $m$ varies in $\cM$.

\begin{definition}
The functions $c_k^G(f)\in \cC^\infty(\cM)$ are called the {\em
Hessian functions} of $f$ with respect to $G$.
\end{definition}

\noindent Let:
\be
\sigma_k(z_1,\ldots, z_n)\eqdef \sum_{1\leq i_1<\ldots<i_k\leq n} z_{i_1}\ldots z_{i_k}\in \R[z_1,\ldots, z_n]~~
\ee
be the elementary symmetric polynomials in $n$ variables, where
$k$ runs from $0$ to $n$. We have:
\be
c_k^G(f)(m)=\sigma_k(\lambda_1(f)(m),\ldots, \lambda_n(f)(m))~,~\forall m\in \cM~~,
\ee
where $\lambda_j(f)$ are functions given by the real eigenvalues of
the $G$-symmetric endomorphism $\hHess_G(f)$ of $T\cM$. Let $\wedge^k
\hHess_G(f)\in \End_\R(\wedge^k T\cM)$ be the $k$-exterior power of
this endomorphism. The relations:
\be
c_k(f)=\tr\left[\wedge^k \hHess_G(f)\right]~,~\forall k=0,\ldots, n
\ee
show that the correspondence $f\rightarrow c_k^G(f)$ gives a
differential operator:
\be
c_k^G:\cC^\infty(\cM)\rightarrow \cC^\infty(\cM)~~
\ee
of order $2k$ (which is non-linear for $k>1$).

\begin{definition}
The differential operator $c_k^G$ is called the {\em $k$-th invariant
  Hessian operator} of $(\cM,G)$.
\end{definition}

\noindent In particular, we have:
\be
c_0=1~~,~~c_1=\tr \left[\hHess_G(\Lambda)\right]=-\Delta_G \Lambda~~,~~c_n=\det \left[\hHess_G(\Lambda)\right]=\rM_G(\Lambda)~~,
\ee
where $\Delta_G=-\div_G\grad_G$ and $\rM_G$ are respectively
the positive Laplacian and the Monge-Amp\`ere operators of $(\cM,G)$.

\begin{definition}
A {\em Hessian equation} on $(\cM,G)$ is a PDE
of the form:
\be
F\circ (f\times c_1^G(f)\times \ldots \times c_n^G(f))=0~~,
\ee
where $F\in \cC^\infty(\R\times \cM)$ is given and the unknown $f$ is
a smooth real-valued function defined on $\cM$.
\end{definition}

\noindent We refer the reader to \cite{hessian1,hessian2} for background on Hessian equations. 

\begin{prop}{\rm \cite{Hesse}}
\label{prop:HessianSystem}
The Hesse equation \eqref{HessCond} is equivalent with the following
system of Hessian equations:
\be
c_k^G(\Lambda)= \frac{n!}{k! (n-k)!}\Lambda^k~~,~~\forall k=1,\ldots, n~~. 
\ee
In particular, any Hesse function $\Lambda$ satisfies the Helmholtz equation
$\Delta_G\Lambda=-n\Lambda$ and the Monge-Amp\`ere equation
$\rM_G(\Lambda)=\Lambda^n$.
\end{prop}

\noindent Since the right hand side of the Helmholtz equation has the
``wrong sign'' for the positive Laplacian $\Delta_G$, this implies:

\begin{cor}
Let $(\cM,G)$ be a Hesse manifold. Then $(\cM,G)$ is non-compact.
\end{cor}

\subsection{The space of Hesse functions}

The space of Hesse functions of any Riemannian manifold is finite-dimensional.
More precisely:

\begin{prop} {\rm \cite{Hesse}}
For any Riemannian $n$-manifold $(\cM,G)$, we have $\fh_G(\cM)\leq n+1$. 
\end{prop}

\noindent The space of Hesse functions carries a natural symmetric
bilinear pairing which is invariant under the action of the isometry
group. We start by defining a certain extension of this pairing.

\begin{definition}
The {\em extended Hesse pairing} of $(\cM,G)$ is the symmetric $\R$-bilinear
map $(~,~)^e_G:\cC^\infty(\cM)\times \cC^\infty(\cM)\rightarrow
\cC^\infty(\cM)$ defined through:
\be
(f_1,f_2)^e_G\eqdef f_1 f_2 -\langle \dd f_1, \dd f_2 \rangle_G=f_1 f_2-
\langle \grad_G f_1,\grad_G f_2 \rangle_G~.
\ee
\end{definition}

\noindent Recall that we assume $\cM$ to be connected. An easy
computation using the Hesse equation gives:

\begin{prop} {\rm \cite{Hesse}}
\label{prop:constant}
The function $(\Lambda_1,\Lambda_2)^e_G$ is constant on $\cM$ for any
Hesse functions $\Lambda_1,\Lambda_2\in \cH_G(\cM)$. Hence the
restriction of the extended Hesse pairing to the subspace
$\cH_G(\cM)\subset \cC^\infty(\cM)$ gives an $\R$-valued bilinear
pairing:
\be
(~,~)_G:\cH_G(\cM)\times \cH_G(\cM)\rightarrow\R~~,
\ee
which we shall call the {\em Hesse pairing} of $(\cM,G)$. 
\end{prop}

\begin{rmk}
By Proposition \ref{prop:constant}, any Hesse function $\Lambda\in
\cH_G(\cM)$ satisfies the nonlinear first order ODE:
\ben
\label{geneik}
||\grad_G\Lambda||_G^2=\Lambda^2-(\Lambda,\Lambda)_G~~,
\een
where $(\Lambda,\Lambda)_G$ is a constant. Notice that
$||\grad_G\Lambda||_G^2=||\dd \Lambda||^2_G$ .  When
$(\Lambda,\Lambda)_G=0$, equation \eqref{geneik} reduces on the
complement of the zero locus of $\Lambda$ to the eikonal equation of
$(\cM,G)$ for the function $f\eqdef \log|\Lambda|$:
\ben
\label{eik}
||\grad_Gf||_G^2=1~~.
\een
Hence \eqref{geneik} can be viewed as a generalization of the eikonal
equation.
\end{rmk}

\begin{definition}
The {\em Hesse norm} of a Hesse function $\Lambda$ is the
non-negative number $\kappa_\Lambda\eqdef
\sqrt{|(\Lambda,\Lambda)_G|}$, while its {\em type indicator}
is the sign factor $\epsilon_\Lambda\eqdef\sign(\Lambda,\Lambda)_G$.  A
non-trivial Hesse function $\Lambda$ is called {\em timelike}, {\em
  spacelike} or {\em lightlike} when
$\epsilon_\Lambda$ equals $+1$, $-1$ or $0$ respectively.
\end{definition}

\noindent Notice that lightlike Hesse functions form a cone in
$\cH_G(\cM)$.

\subsection{The Morse property of Hesse functions}

\begin{prop}{\rm \cite{Hesse}}
\label{prop:Morse}
Let $\Lambda\in \cH_G(\cM)$ be a nontrivial Hesse function. Then
$\Lambda$ has isolated critical points, i.e. it is a Morse function on
$\cM$. Moreover, the following statements hold:
\begin{itemize}
\itemsep 0.0em
\item If $\Lambda$ is timelike, then $\Lambda$ does not have any
  zeroes on $\cM$.
\item If $\Lambda$ is spacelike, then $\Lambda$ does not have any
  critical points on $\cM$.
\item If $\Lambda$ is lightlike, then $\Lambda$ has neither zeroes
  nor critical points on $\cM$.
\end{itemize}
Hence $\Lambda$ can have zeroes iff $(\Lambda,\Lambda)_G<0$ and it can
have critical points iff $(\Lambda,\Lambda)_G>0$.
\end{prop}

\section{The gradient flow of Hesse functions}

\noindent Let $\Lambda\in \cH_G(\cM)$ be a non-trivial Hesse function
and consider the gradient flow equation:
\ben
\label{gradeq}
\gamma'(q)=-(\grad_\cG \Lambda)(\gamma(q))
\een
for smooth curves $\gamma:I\rightarrow \cM$, where $I$ is an interval
and $\gamma'(q)\eqdef \frac{\dd \gamma}{\dd q}$. This equation fixes
the parameter $q$ of a solution $\gamma$ (which we shall call the {\em
gradient flow parameter}) up to translation by a constant. The {\em
level set parameter} $\lambda$ of $\gamma$ is defined through:
\be
\lambda(q)\eqdef \Lambda(\gamma(q))
\ee
and decreases as the gradient flow parameter increases.

\begin{prop}{\rm \cite{Hesse}}
\label{prop:levelparam}
The level set and gradient flow parameters of any gradient flow curve
$\gamma$ of $\Lambda$ satisfy:
\ben
\label{ql}
\dd q=-\frac{\dd \lambda}{||\dd_{\gamma(\lambda)}\Lambda||_G^2}=
\frac{\dd \lambda}{(\Lambda,\Lambda)_G-\lambda^2}
\een
and are related through:
\ben
\label{qlambdarel}
q=\threepartdef{\frac{1}{\kappa_\Lambda}\arctanh\left(\frac{\lambda-\lambda_0}{\kappa_\Lambda}\right)~~,}
{~\epsilon_\Lambda=+1}{-\frac{1}{\kappa_\Lambda}\arctan\left(\frac{\lambda-\lambda_0}{\kappa_\Lambda}\right)~,}
{~\epsilon_\Lambda=-1}{\frac{\lambda_0}{\lambda}~~~~~~~~~~~~~~~~~~,}{~\epsilon_\Lambda=~0}~~.
\een
where $\lambda_0$ is an integration constant and:
\ben
\label{lambdaqrel}
\lambda=\threepartdef{\kappa_\Lambda\tanh(\kappa_\Lambda q)~~,}{~\epsilon_\Lambda=+1}{-\kappa \tan(\kappa_\Lambda q)~~,}{~\epsilon_\Lambda=-1}{\frac{1}{q}~~,}{~\epsilon_\Lambda=0}~~,
\een
where $\kappa_\Lambda$ and $\epsilon_\Lambda$ are the Hesse norm and
type indicator of $\lambda$. In the formulas above, we chose the
integration constant $\lambda_0$ such that $q|_{\lambda=\lambda_0}=0$
when $\epsilon_\Lambda=\{-1,+1\}$ and $q|_{\lambda=\lambda_0}=1$ when
$\epsilon_\Lambda=0$.
\end{prop}

\subsection{The general form of Hesse functions}

The relation to the eikonal equation allows us to express Hesse
functions using the distance function of the Riemannian manifold
$(\cM,G)$, for whose properties we refer the reader to
\cite{Petersen}. We need a few preparations before stating this result.

\begin{prop}{\rm \cite{Hesse}}
Suppose that the Riemannian manifold $(\cM,G)$ is complete and
let $\Lambda\in \cH_G(\cM)\setminus \{0\}$ be a non-trivial Hesse
function. Then the following statements hold:
\begin{enumerate}
\itemsep 0.0em
\item If $\Lambda$ is timelike, then the vanishing locus $Z(\Lambda)$
of $\Lambda$ is empty and hence $\Lambda$ has constant sign (denoted
$\eta_\Lambda$) on $\cM$. Moreover, $\Lambda$ has exactly one critical
point, with critical value $\eta_\Lambda\kappa_\Lambda$, which is a
global minimum or maximum according to whether $\eta_\Lambda=+1$ or
$-1$.
\item If $\Lambda$ is spacelike, then the set $\Crit(\Lambda)$ of
critical points of $\Lambda$ is empty. Moreover, the vanishing locus
of $\Lambda$ coincides with the $\kappa_\Lambda$-level set of
the function $||\dd\Lambda||_G$:
\be
\label{Zspacelike}  
Z(\Lambda)=\{m\in \cM \, | \, ||\dd_m\Lambda||_G=\kappa_\Lambda\}~~,
\ee
which is a non-singular hypersurface in $\cM$.
\item If $\Lambda$ is lightlike, then
$Z(\Lambda)=\Crit(\Lambda)=\emptyset$ and hence $\Lambda$ has constant
sign on $\cM$, which we denote by $\eta_\Lambda$.
\end{enumerate}
\end{prop}

\begin{definition}
Suppose that $(\cM,G)$ is complete. Then a timelike or lightlike
non-trivial Hesse function $\Lambda\in \cH_G(\cM)\setminus\{0\}$ is
called {\em future (resp. past) pointing} when $\eta_\Lambda=+1$
(resp. $-1$).
\end{definition}

\begin{definition}
Let $\Lambda\in \cH_G(\cM)\setminus \{0\}$ be a non-trivial Hesse
function of $(\cM,G)$. The {\em characteristic set} of $\Lambda$ is
the following closed subset of $\cM$:
\be
\label{Q}
Q_\Lambda\eqdef \threepartdef{\Crit(\Lambda)~,~}{~\Lambda~{\rm~is~timelike}}{Z(\Lambda)~,~}{~\Lambda~{\rm is~spacelike}}{\cM_{|\Lambda|}(1)~,~}{~\Lambda~{\rm~is~lightlike}}~.
\ee
The {\em characteristic constant} of $\Lambda$ is defined through:
\be
\label{C}
C_\Lambda\eqdef \threepartdef{\kappa_\Lambda~,~}{~\epsilon_\Lambda=+1}{0~,~}{~\epsilon_\Lambda=-1}{1~,~}{~\epsilon_\Lambda=0}~.
\ee
\end{definition}

\noindent Set $\cU_\Lambda\eqdef \cM\setminus \Crit(\Lambda)$. We have:
\be
Q_\Lambda=\{m\in \cU_\Lambda ~\, | \, |\Lambda(m)|=C_\Lambda\}~~.
\ee

\begin{definition}
Let $\Lambda\in \cH_G(\cM)\setminus \{0\}$ be a non-trivial Hesse
function of $\cM$. The {\em characteristic sign function} of
$\Lambda$ is the function $\Theta_\Lambda:\cM\rightarrow \R$ defined
through:
\ben
\label{Theta}
\Theta_\Lambda(m)\eqdef \threepartdef{1~,~}{~\epsilon_\Lambda=+1}{\sign(\Lambda(m))~,~}{~\epsilon_\Lambda=-1}{\sign(|\Lambda(m)|-1)~,~}{~\epsilon_\Lambda=0}~~.
\een
The {\em $\Lambda$-distance function} of $(\cM,G)$ is the function
$d_\Lambda:\cM\rightarrow \R$ defined through:
\ben
\label{d}
d_\Lambda(m)\eqdef \Theta_\Lambda(m) \dist_G(m,Q_\Lambda)~~.
\een
\end{definition}

\begin{thm}{\cite{Hesse}}
Let $\Lambda\in \cH_G(\cM)$ be a non-trivial Hesse function. Then the
following relation holds for all $m\in \cM$:
\be
\Lambda(m)=\threepartdef{\sign(\Lambda)\kappa_\Lambda\cosh d_\Lambda(m) ~,~}
{~\epsilon_\Lambda=+1}{\kappa_\Lambda\sinh d_\Lambda(m) ~,~}
{~\epsilon_\Lambda=-1}{\sign(\Lambda) e^{d_\Lambda(m)}~~~~~~~~~~~~,~}{~\epsilon_\Lambda=~0}~~.
\ee
\end{thm}

\subsection{Maximally Hesse manifolds are Poincar\'e balls}

Complete Hesse manifolds of maximal Hesse index turn out to be particularly simple, namely
any such manifold is isometric with a Poincar\'e ball.

\begin{definition}
A Hesse manifold $(\cM,G)$ is called {\em maximally Hesse} if
$\fh_G(\cM)=n+1$.
\end{definition}

\noindent Recall that a Riemannian manifold $(\cM,G)$ is {\em
hyperbolic} if its metric $G$ has unit negative sectional
curvature. Up to isometry, there exists a unique simply connected and
complete hyperbolic $n$-manifold, namely the Poincar\'e $n$-ball,
whose description we recall below. Let:
\be
\rD^n\!\eqdef \!\{u \in \R^n | 0\leq ||u||_E \!<\!1 \}
\ee
be the open unit $n$-ball, where $||~||_E$ is the
Euclidean norm on $\R^n$. The {\em Poincar\'e ball metric} is the
complete Riemannian metric $G_n$ on $\rD^n$ whose squared line element
is given by:
\be
\dd s_{G_n}^2=\frac{4}{(1-||u||_E^2)^2}\sum_{i=1}^n (\dd u^i)^2~~.
\ee
The $n$-dimensional {\em Poincar\'e ball} is the complete hyperbolic
manifold $\mD^n\eqdef (\rD^n,G_n)$.

\begin{prop}{\rm \cite{Hesse}}
A complete Riemannian $n$-manifold $(\cM,G)$ is maximally Hesse iff it is
isometric to the Poincar\'e ball $\mD^n$.
\end{prop}

\noindent The space of Hesse functions of $\mD^n$ identifies naturally with a Minkowski
space, as we explain next. Consider the $(n+1)$-dimensional Minkowski space $\R^{1,n}\eqdef
(\R^{n+1},(~,~))$ where:
\ben
\label{Mink0}
(X,Y)\eqdef X^0Y^0-\sum_{i=1}^n {X^i Y^i}=\eta^{\mu\nu}X^\mu Y^\nu
\een
is the Minkowski pairing. We denote the canonical basis of $\R^{n+1}$ by:
\be
E_0=(1,0,0,\ldots, 0)~~,~~E_1=(0,1,0,\ldots, 0)~~,~~\ldots~~,~~E_n=(0,0,0,\ldots, 1)~~.
\ee
Let $\vec{X}\eqdef (X^1,\ldots, X^n)$, so that $X=(X^0,\vec{X})$ and:
\be
(X,X)=X^0Y^0 -\vec{X}\cdot \vec{Y}~~,
\ee
where $\cdot$ denotes the Euclidean scalar product in $\R^n$.  Let
$\rS_n^+$ be the future sheet of the hyperboloid defined by the
equation $(X,X)=1$:
\be
\rS_n^+\eqdef \{X\in \R^{n+1}| (X,X)=1~~\&~~ X^0>0\}=\Big{\{}X\in \R^{n+1}| X^0=\sqrt{1+||\vec{X}||_E^2} \Big{\}}~~.
\ee
Then $\rS_n^+$ is diffeomorphic with $\rD^n$ through the {\em Weierstrass map} $\Xi:\rD^n\rightarrow \rS_n^+$, which is defined
through:
\ben
\label{Xi0}
\Xi(u)\eqdef \left(\frac{1+||u||_E^2}{1-||u||_E^2}, \frac{2 u}{1-||u||_E^2}\right)~,~\forall u\in \rD^n
\een
and whose inverse $\Xi^{-1}:\rS_n^+\rightarrow \mD^n$ is given by:
\be
\Xi^{-1}(X)=\frac{\vec{X}}{X^0+1}=\frac{\vec{X}}{1+\sqrt{1+||\vec{X}||_E^2}}~,~\forall X\in \rS_n^+~~.
\ee
Notice the relations
\be
||u||_E^2=\frac{\Xi^0(u)-1}{X^0(u)+1}~~\Longleftrightarrow \Xi^0(u)=\frac{1+||u||_E^2}{1-||u||_E^2}~~.
\ee
The components $\Xi^\mu(u)$ (which satisfy the relation
$\eta_{\mu\nu} \Xi^\mu (u) \Xi^\nu(u)=-1$) are the classical
{\em Weierstrass coordinates} of the point $u\in \rD^n$. The
Weierstrass map can be viewed as the projection of $\rD^n$ onto $\rS_n^+$
from the point $-E_0=(-1,0,\ldots, 0)$ of $\R^{1,n}$. It is well-known
that $\Xi$ is an isometry from $\mD^n$ to $\rS_n^+$ when $\rS_n^+$ is
endowed with the Riemannian metric induced by the {\em opposite} of
the Minkowski metric \eqref{Mink0}. We can now state the result
announced above:

\begin{thm}{\rm \cite{Hesse}}
\label{thm:Hesse}
For any $n>1$, there exists a bijective isometry $\bLambda:
\R^{1,n}\stackrel{\sim}{\rightarrow} (\cH_{G_n}(\rD^n),(~,~)_{G_n})$ such
that:
\be
\bLambda(E_\mu)=\Xi_\mu\eqdef \eta_{\mu\nu} \Xi^\nu~~,~~\forall \mu\in \{0,\ldots, n\}~~.
\ee
\end{thm}

\subsection{The Hesse sheaf and local Hesse index}

Let $(\cM,G)$ be a Riemannian $n$-manifold. The Hesse equation
naturally defines a sheaf of vector spaces on $\cM$.

\begin{definition}
A {\em local Hesse function} of $\cM$ relative to
$G$ is a locally defined solution of the Hesse equation of
$(\cM,G)$. The {\em Hesse sheaf} of $(\cM,G)$ is the sheaf of local
Hesse functions of $(\cM,G)$.
\end{definition}

\begin{prop}{\rm \cite{Hesse}}
We have $\rk \cH_G\leq n+1$.
\end{prop}

\begin{definition}
We say $(\cM,G)$ is {\em locally Hesse} if its Hesse sheaf does not
vanish, i.e. if $\rk\cH_G>0$.
\end{definition}

\noindent Notice that $\cH_G(\cM)=H^0(\cH_G)$ and hence
$\fh_G(\cM)=h^0(\cH_G)=\dim_\R H^0(\cH_G)$. Thus $(\cM,G)$
is globally Hesse iff its Hesse sheaf admits nontrivial global
sections.

\subsection{Locally maximally Hesse manifolds are elementary hyperbolic space forms}

\begin{definition}
A Riemannian manifold $(\cM,\cH)$ is called {\em locally maximally
Hesse} if $\rk\cH_G=n+1$.
\end{definition}

\begin{thm}{\rm \cite{Hesse}}
A Riemannian manifold is locally maximally Hesse iff it is hyperbolic.
\end{thm}

\noindent Note that a general hyperbolic manifold need not be Hesse. The
situation is clarified by the following result.

\begin{prop}{\rm \cite{Hesse}}
Let $(\cM,G)$ be a complete Riemannian manifold. The following are
equivalent:
\begin{itemize} \itemsep 0.0em
\item $(\cM,G)$ is hyperbolic and globally Hesse.
\item $(\cM,G)$ is an elementary hyperbolic space form.
\end{itemize}
In this case, $(\cM,G)$ is maximally Hesse iff it is
isometric with a Poincar\'e ball.
\end{prop}

\noindent Hyperbolic uniformization and the notion of elementary hyperbolic space form are
recalled in Appendix \ref{app:hyp}.

\section{Acknowledgements}
This work was partly supported by grant PN 19060101/2019-2022 and partly by IBS-R003-D1. 

\appendix

\section{Hyperbolic uniformization and elementary hyperbolic space forms}
\label{app:hyp}

Recall that the group of orientation-preserving isometries of the
Poincar\'e $n$-ball is naturally isomorphic with the connected
component $\SO_0(1,n)$ of the identity in the Lorentz group
$\SO(1,n)$. Indeed, $\SO_0(1,n)$ acts linearly on $\R^{n+1}$ (and
hence on the hyperboloid model $\rS_n^+$ of $\mD^n$) through the
fundamental representation $R:\SO_0(1,n)\rightarrow \Aut_\R(\R^{n+1})$:
\be
R_A(x)= A X~~,~~\forall A\in \SO_0(1,n)~~,~~\forall X\in \R^{n+1}~~,
\ee
where $R_A\eqdef R(A)$. Since this action preserves orientation as
well as the Minkowski pairing (and hence the Riemannian metric induced
on $\rS_n^+$), it induces a morphism of groups
$\psi:\SO_0(1,n)\rightarrow \Iso_+(\mD^n)$, which turns out to be an
isomorphism.  For any $A\in \SO_0(1,n)$, the corresponding isometry
$\psi_A\eqdef \psi(A)$ of the Poincar\'e ball is determined uniquely
by the following condition, which encodes $\SO_0(1,n)$-equivariance of
the Weierstrass map:
\ben
\label{XiA}
\Xi\circ \psi=R\circ \Xi~~,~~\mathrm{i.e.}~~\Xi(\psi_A(\vec{u}))=A\Xi(\vec{u})~,~~\forall A\in \SO_o(1,n)~~,~~\forall \vec{u}\in \rD^n~~.
\een
A general element $A\in \SO_0(1,n)$ has the form: 
\be
A(\vec{v})=\left[\begin{array}{cc} \gamma~~&~~~~~ -\gamma(\vec{v}) \vec{v}\\
-\gamma(\vec{v}) \vec{v\,}^T~~&~~~ I_n+(\gamma(\vec{v})-1) \hv\otimes \hv\end{array} \right]
\ee
where $\vec{v}\in \R^n$ and we defined: 
\be
v\eqdef ||\vec{v}||_E~~,~~\hv\eqdef \frac{\vec{v}}{v}~~, ~~\gamma(\vec{v})\eqdef
\frac{1}{\sqrt{1-v^2}}~~,~~\hv\otimes \hv=(\hat{v}_i
\hat{v}_j)_{i,j=1,\ldots n}\!=\!\left(\frac{v_i v_j}{v^2}\right)_{i,j=1,\ldots n}.
\ee

\noindent The following result is classical:

\begin{prop}
For any $\vec{v}\in \R^n$, $\vu\in \rD^n$ and $A\in \SO_0(1,n)$, we
have:
\ben
\label{psiA}
\psi_{A(\vec{v})}(\vu)=\frac{2u+2(\gamma(\vec{v})-1) (\hv\cdot \vec{u})\hv -\gamma(\vec{v}) (1+||\vec{u}||_E^2) \vec{v}}{1-||\vec{u}||_E^2+\gamma(\vec{v})(1+||\vec{u}||_E^2- 2\vec{v}\cdot \vec{u})}~~.
\een
\end{prop}

By the uniformization theorem of hyperbolic geometry (see
\cite{Ratcliffe}), any oriented and complete hyperbolic $n$-manifold
$(\cM,G)$ can be written as the Riemannian quotient of the unit
hyperbolic ball $\mD^n$ through a discrete subgroup $\Gamma\in
\Iso(\mD^n)\simeq \SO_0(1,n)$ called the {\em uniformizing group} of
$(\cM,G)$. Notice that $\Gamma$ is isomorphic with the fundamental
group of $\cM$. We remind the reader of the following classical
notions, for which we refer him or her to \cite{Ratcliffe}.

\begin{definition}
A discrete subgroup $\Gamma$ of $\SO_0(1,n)$ is called {\em
elementary} if its action on the closure of the Poincar\'e ball
fixes at least one point.
\end{definition}

\begin{definition}
An $n$-dimensional {\em elementary hyperbolic space
form} is a complete hyperbolic $n$-manifold uniformized by a
torsion-free elementary discrete subgroup $\rGamma\subset
\SO_0(1,n)$.
\end{definition}

\noindent A torsion-free elementary discrete subgroup $\Gamma\subset
\SO_0(1,n)$ is called:
\begin{itemize}
\itemsep 0.0em
\item {\em elliptic}, if it conjugates to a subgroup of the {\em
canonical rotation group} $ \cR_n\eqdef \Stab_{\SO_0(1,n)}(E_0)\simeq
\SO(n)$. In this case, $\rGamma$ is finite.
\item {\em hyperbolic}, if it conjugates to a subgroup of the {\em
canonical squeeze group} $\cT_n\eqdef \Stab_{\SO_0(1,n)}(E_n)\simeq
\SO(1,n-1)$.  In this case, $\rGamma$ is a hyperbolic cyclic group.
\item {\em parabolic}, if it conjugates to a subgroup of the {\em
canonical shear group} $\cP_n\eqdef \Stab_{\SO_0(1,n)}(E_0+E_n)\simeq
\ISO(n)$. In this case, $\rGamma$ is a free Abelian group of rank at
most $n-1$.
\end{itemize}

\noindent Any nontrivial torsion-free elementary discrete subgroup of
$\SO_0(1,n)$ is either elliptic, parabolic or hyperbolic, while the trivial
subgroup of $\SO_0(1,n)$ belongs to each of these classes. An
elementary hyperbolic space form different from $\mD^n$ is called
elliptic, parabolic or hyperbolic if its uniformizing group $\Gamma$
is of that type.

\end{document}